\documentclass[journal=jctcce,manuscript=article,layout=twocolumn]{achemso}
\usepackage[version=3]{mhchem} 
\usepackage{xcolor}
\usepackage{mycommands}
\usepackage{bbold}
\usepackage[symbol]{footmisc}

\usepackage{bm}
\renewcommand{\vec}[1]{\bm{#1}}

\author{J. Amira Geuther}
\author{Kasra Asnaashari}
\author{Jeremy O. Richardson}
\email{jeremy.richardson@phys.chem.ethz.ch}
\affiliation[ETH Zurich]
{Department of Chemistry and Applied Biosciences, ETH Zurich, 8093 Zurich, Switzerland}

\title[]
  {Time-reversible implementation of MASH for efficient nonadiabatic molecular dynamics}

\abbreviations{}
\begin{document}


\begin{abstract}
In this work, we describe various improved implementations of the mapping approach to surface hopping (MASH) for simulating nonadiabatic dynamics.
These include time-reversible and piecewise-continuous integrators, which is only formally possible because of the deterministic nature of the underlying MASH equations of motion. 
The new algorithms allow for the use of either wave-function overlaps or nonadiabatic coupling vectors to propagate the spin, which encodes the electronic state. 
For a given time-step, $\Delta t$, it is demonstrated that the global error for these methods is $\mathcal{O}(\Delta t^2)$ compared to the $\mathcal{O}(\Delta t)$ error of standard implementations. This allows larger time-steps to be used for a desired error tolerance, or conversely, more accurate observables given a fixed value of $\Delta t$.
The newly developed integrators thus provide further advantages for the MASH method, demonstrating that it can be implemented more efficiently than other surface-hopping approaches, which cannot construct time-reversible integrators due to their stochastic nature.

\end{abstract}

\section{Introduction}
Molecular dynamics (MD) simulates the behaviour of atoms and molecules, providing insights into the behaviour of matter at the atomic level, which is crucial for understanding chemical reactions, material properties, and biological processes.\cite{AllenTildesley}
A fundamental concept of standard MD is the Born--Oppenheimer approximation, \cite{born_zur_1927}  which assumes that the electronic and nuclear motion can be separated by evolving the electronic wave function adiabatically along the nuclear trajectory. 
However, this approximation breaks down in situations where the potential energy surfaces of different electronic states come close to each other or intersect, leading to strong coupling between electronic and nuclear degrees of freedom. These scenarios, known as nonadiabatic processes, are essential in understanding and describing photochemical reactions, electron-transfer processes, and the behaviour of excited states in materials \cite{worth_beyond_2004}.

Unfortunately, solving the full time-dependent Schrödinger equation is not possible for most systems of interest and the computational cost for fully quantum-mechanical methods increases exponentially with the size and complexity of the system studied. 
Thus, several mixed quantum--classical methods have been developed to describe nonadiabatic processes based on independent trajectories, such as various mapping approaches\cite{Meyer1979nonadiabatic,Stock1997mapping} including spin-mapping\cite{spinmap,multispin}.
However, the most commonly used approach in molecular photochemistry is fewest-switches surface hopping (FSSH) \cite{tully_molecular_1990}. This method propagates the classical nuclei on a single adiabatic energy surface while stochastic hops allow for a switch between different electronic states.
It correctly describes certain effects in nonadiabatic processes such as wave-packet branching\cite{barbatti_nonadiabatic_2011}, although it also has several known problems such as an inconsistency with its wave-function coefficients. \cite{ subotnik_understanding_2016,tully_molecular_1990}

Recently, we have developed the mapping approach to surface hopping (MASH) \cite{mannouch_mapping_2023,richardson_nonadiabatic_2025}. It combines concepts from both spin-mapping and surface hopping but shows distinct advantages over those methods in terms of rigour and accuracy at a comparable computational cost.
In particular, unlike FSSH\cite{Landry2011hopping,Landry2012hopping}, MASH captures the correct rate in the Marcus-theory limit without requiring complicated decoherence corrections, \cite{MASHrates}
and it appears to outperform all forms of FSSH (especially those which use the wrong form of momentum rescaling) and even AIMS (ab initio multiple spawning)\cite{Curchod2018review} in first-principles simulations of photochemistry \cite{Mannouch2024MASH}. One argument for why
MASH shows this good behaviour, is because it can be rigorously derived from a short-time limit of the quantum--classical Liouville equation, as we have discussed in previous work. \cite{mannouch_mapping_2023,richardson_nonadiabatic_2025}

This work does not focus on the theoretical accuracy of MASH but rather on the numerical accuracy of its implementation.
In particular, we construct time-reversible integrators, explore different methods to compute the time-derivative couplings and implement variable time-stepping. Note that, this is only possible due to the deterministic nature of the theory, which yields reversible equations of motions that allow one to ``undo" propagation steps as well as to reverse the dynamics. 
Such properties do not hold for FSSH, as this does not have time-reversible equations of motion, and it is also not easy to ``undo'' a propagation step once one has drawn a random number without biasing the stochastic process. 

We will begin with a brief explanation of the MASH method.
Following this, we describe various ways to propagate the spin vector. These are combined with standard methods for nuclear propagation to construct time-reversible and piecewise-continuous algorithms. Finally, we present results from tests on model systems to compare the efficiency of the new algorithms to the previously used non-reversible MASH algorithms, which were based on standard approaches used for FSSH.

\section{Mapping approach to surface hopping}
As the original MASH approach is only applicable to two-level systems, we focus on this case.  However, our integrators can easily be generalized to treat the multistate MASH methods so far proposed.\cite{Runeson2023MASH,unSMASH}

In many ways, MASH is similar to FSSH: the nuclei are evolved on a single active adiabatic potential energy surface and can switch surface by jumping to a different state. While the nuclei are treated classically with coordinates $\vec q$ and associated momentum $\vec p$, the electronic state is treated quantum mechanically, either using wave-function coefficients, or equivalently  by a spin vector $\vec S=(S_x,S_y,S_z)$ on the Bloch sphere.
In MASH, rather than using a stochastic process, the active state is chosen deterministically according to the instantaneous position of the spin vector.
If the spin is in the southern hemisphere of the Bloch sphere, the nuclei should be evolved on the lower adiabatic energy surface, $V_0(\vec q)$; 
if it is in the northern hemisphere, they are propagated on the upper surface, $V_1(\vec q)$. 
The two states interact via the nonadiabatic coupling vector, $\vec d(\vec q)$, defined by the elements
     \begin{equation}\label{NAC}
         d_j(\vec q) = \Braket{ \Phi_1(\bm{q}) | \frac{\partial}{\partial q_j} | \Phi_0(\bm{q})}
     \end{equation}
     where $\Phi_0$ and $\Phi_1$ are the stationary, real-valued electronic wave functions of the two adiabatic states.

The equations of motion are (with $\hbar=1$) \cite{mannouch_mapping_2023}
\begin{subequations}
     \begin{align}
     \dot{q}_j &= \frac{p_j}{m_j}
     \\
       \dot{p}_j
       &= -\pder{V_0}{q_j}h(-S_z) - \pder{V_1}{q_j}h(S_z) \nonumber\\&\quad+
       4V_z (\vec q)d_j(\vec q)S_x\delta(S_z)
     \\
     \dot{\vec S} &= \bigg(0, 2\sum_j \frac{p_jd_j(\vec q)}{m_j}, 2V_z(\vec q)\bigg) \times \vec{S}
     \label{S_EOM}
     \end{align}
\end{subequations}
     where $m_j$ is the mass associated with the $j$th degree of freedom and $V_z = \frac{1}{2}(V_1 - V_0)$. 
     The Heaviside step function, $h(S_z)$, determines the active surface and hence the force felt by the nuclei.
     When $S_z$ passes through zero (the Bloch sphere's equator), the electrons attempt to hop to the other state and the nuclei feel an impulse in the direction of the nonadiabatic coupling vector $\vec d(\vec q)$ due to the the Dirac delta function, $\delta(S_z)$. This ensures that the energy
     \begin{align}
\mathcal{E}( \vec q,\vec p,\vec  S)&= \sum_j \frac{p_j^2}{2m_j} + h(-S_z)V_0(\vec q) + h(S_z)V_1(\vec q)
\end{align}
is conserved.
     If not enough energy is available to hop up,
     the attempt is rejected and the momenta are reflected [along the direction of $\vec d(\vec q)$], which was also the suggestion first made by Tully.\cite{HammesSchiffer1994FSSH}

Note that MASH is defined in the kinematic representation\cite{Cotton2017mapping} and its dynamics are not generated by a Hamiltonian. One cannot thus hope to obtain a symplectic integrator (at least not with a canonical structure), as has been achieved for mapping and spin-mapping methods.\cite{Church2018MQCIVR,symplectic}

Unlike other surface-hopping methods \cite{jain_surface_2015}, however, the MASH equations of motion are fully deterministic and time-reversible (as is clear from the replacement $\vec p\mapsto-\vec p,S_y\mapsto-S_y$). The fact that the differential equations are time-reversible inspires us to search for a numerical implementation which retains this symmetry with arbitrary step sizes.

\subsection{Nuclear propagation}

Before tackling the coupled case, we first consider propagating the nuclear and spin vector separately.  In particular, 
it is clear that the nuclei can be evolved using the standard velocity-Verlet algorithm \cite{leimkuhler_simulating_2005} given by 
\begin{subequations}
\begin{align} \label{eq9}
    \vec p(t+\thalf\Delta t) &= \vec p(t) + \frac{\Delta t}{2} \vec F(\vec q(t))
    \\ \vec q(t+\Delta t) &= \vec q(t) + \Delta t \, \, \frac{ \vec p(t+\thalf\Delta t)}{m}
    \\\vec p(t+\Delta t) &=  \vec p(t+\thalf\Delta t) + \frac{\Delta t}{2} \vec F(\vec q(t+\Delta t))
\end{align} 
\end{subequations}
where $\vec F$ is given by 
\begin{equation}
    F_j = - \frac{\partial V_0(\vec q)}{\partial q_j}h(-S_z) - \frac{\partial V_1(\vec q)}{\partial q_j}h(S_z)
\end{equation}

There are a number of ways of propagating $\vec S$, which we discuss later.  In any case, after each update, we test for hops by comparing $\sgn(S_z)$ before and after the step. After a hop, the momentum is rescaled.The rigorous connection between MASH and the QCLE gives a unique procedure for rescaling the mass-weighted momentum ($\tilde{p}_j=p_j/\sqrt{m_j}$) along the direction of the mass-weighted NAC ($\tilde{d}_j=d_j/\sqrt{m_j}$). 
Therefore, the total available energy is given by
\begin{equation}
    E_{d} = \thalf\tilde{p}_{d,\rm init}^2 + V_{\rm init}
\end{equation}
where $\tilde{p}_{d,\rm init} =\tilde{\bm d}\cdot\tilde{\bm p}_{\rm init}/||\tilde{\bm d}||$ and $V_{\rm init}$ are the initial momentum along the NAC and the initial potential before the hop. 
To determine whether there is enough energy available to accept the hop,
the final potential energy, $V_{\rm fin}$, is computed on the new surface.
If $E_{d} > V_{\rm fin}$, the hop is accepted and the momentum is rescaled according to 
\begin{equation}
    \tilde{p}_{d,\rm fin} = {\rm sgn}(\tilde{p}_{d,\rm init})\sqrt{\tilde{p}_{d,\rm init}^2 + 2(V_{\rm init} - V_{\rm fin})}
\end{equation} 
to give 
\begin{equation}
    \tilde{\vec p}_{\rm fin} = \tilde{\vec p}_{\rm init} - \frac{\tilde{\bm d}}{||\tilde{\bm{d}}||}(\tilde{p}_{d,\rm init} - \tilde{p}_{d,\rm fin})
\end{equation} 
If $E_{d} < V_{\rm fin}$, the hop is rejected and the nuclei are reflected by inverting the component of the momentum along the nonadiabatic coupling, giving
\begin{equation}
    \tilde{\vec p}_{\rm fin} = \tilde{\vec p}_{\rm init} - 2\frac{\tilde{\bm d}}{||\tilde{\bm{d}}||}\tilde{p}_{d,\rm init}.
\end{equation}
Finally, each degree of freedom, $j$, is multiplied by $\sqrt{m_j}$ to give the full, rescaled momentum $\vec p_{\rm fin}$.
Additionally, the $S_z$ coordinate is inverted so as to remain in the initial hemisphere.

\subsection{Spin propagation with NACs}

The simplest method to propagate the spin vector according to its equation of motion \eqn{S_EOM} is
\begin{equation} \label{eq20}
    \vec S(t+\Delta t)= \text{e}^{\bm \Omega_\text{NAC}\Delta t} \, \vec S(t)
\end{equation}
with 
\begin{equation}
        \bm{\Omega}_\text{NAC} = \begin{pmatrix}
            0 & -\Delta V & 2T \cr
            \Delta V & 0 & 0 \cr
            -2T & 0 & 0
        \end{pmatrix}
\end{equation}
where
\begin{equation} 
    \Delta V = V_1(\vec q) - V_0(\vec q) = 2V_z(\vec q)
\end{equation}
and
\begin{equation} \label{T_NACs}
    T = \sum_j\frac{d_j(\vec q)p_j}{m_j} 
\end{equation}

This approach is particularly simple to implement (assuming NACs are available) as it only requires information from one electronic-structure calculation at a time. However, the well-known disadvantage of using NACs is that they can be strongly peaked for systems with weak diabatic couplings \cite{meek_evaluation_2014,plasser_surface_2012}. This means that a very short time-step may be necessary to simulate such systems correctly.

\subsection{Spin propagation with averaged time-derivative couplings}

An alternative to directly using NACs, is to reformulate the theory in terms of wave-function overlaps.
The advantage of this is that it will correctly capture the effect of narrow couplings which were passed between $t$ and $t+\Delta t$ without requiring a short time-step \cite{meek_evaluation_2014}.
It does, however, bring an extra complication that knowledge of the wave functions at two points in time are required, which, as we will see later, makes constructing reversible integrators more difficult.

The overlap matrix is defined as
\begin{align}
	O_{nm}(t,t') = \braket{\Phi_n(\bm{q}(t)) | \Phi_m(\bm{q}(t'))}
\end{align}
Following Jain et al.,\cite{Jain2016AFSSH}
we note that
\begin{align}\label{DE}
	\der{}{t'}\bm{O}(t,t') = \bm{O}(t,t')\bm{T}(t')
\end{align}
where the time-derivative coupling matrix has elements
\begin{align}
	T_{nm}(t) = \Braket{\Phi_n|\der{}{t}|\Phi_m} = \Braket{\Phi_n|\pder{}{\bm{q}}|\Phi_m} \cdot \der{\bm{q}}{t}
\end{align}
In the two-state case 
these can be written as\footnote[1]{In an ab initio simulation, where there are formally an infinite number of states truncated to a set of two, it is necessary to apply Löwdin orthogonalization in order to achieve the required form.}
\begin{align}\label{ovl}
	\bm{O}(t,t+\Delta t) &= \begin{pmatrix} \cos\frac{\chi}{2} & \sin\frac{\chi}{2} \\ -\sin\frac{\chi}{2} & \cos\frac{\chi}{2} \end{pmatrix}
	= \exp\left(\iu\hat{\sigma}_y \frac{\chi}{2}\right)
\end{align}
where $\hat{\sigma}_y$ is a Pauli matrix
and
\begin{align}
	\bm{T}(t) = -\iu\hat{\sigma}_y T(t) = \begin{pmatrix} 0 & -T(t) \\ T(t) & 0 \end{pmatrix}
\end{align}
where $T(t)$ was defined in \eqn{T_NACs}.
It is necessary to impose a sign convention such that the wave functions change smoothly along the trajectory.  We choose the arbitrary global phase (or the sign in our real-valued case) of each wave function such that it has a positive overlap with  the corresponding state at the previous step.
This ensures that the diagonal elements of the overlap matrix are positive and therefore $-\pi\le\chi\le\pi$.

The solution of Eq. \ref{DE} is
\begin{subequations}
\begin{align}
	\bm{O}(t,t+\Delta t) &= \mathcal{T} \exp\left(\int_t^{t+\Delta t} \bm{T}(t') \, \rmd t'\right)
	\\ &= \exp\left(-\iu \hat{\sigma}_y \overline{T} \Delta t\right)
\end{align}
\end{subequations}
where the time-ordering operator, $\mathcal{T}$, is not required in the two-state case as $\bm{T}(t')$ commutes with itself at different times.
We are thus naturally led to defining the averaged time-derivative coupling as
\begin{align}
	\overline{T} = \frac{1}{\Delta t} \int_t^{t+\Delta t} T(t') \, \rmd t'
\end{align}

Putting all this together, 
we obtain
\begin{align} \label{Tbar}
	\overline{T}\Delta t = -\frac{\chi}{2} = \arcsin(O_{10})
\end{align}
which is the formula used in practical applications to obtain the averaged time-derivative coupling directly from the overlap matrix.

To propagate the spin vector, we write
\begin{subequations}
\begin{align}\label{spin ATDC}
    \bm{S}(t+\Delta t) &= \mathcal{T} \eu{\int_t^{t+\Delta t} \bm{\Omega}_\text{NAC}(t') \,\rmd t'} \, \bm{S}(t)
    \\ &\approx \eu{\int_t^{t+\Delta t} \bm{\Omega}_\text{NAC}(t') \,\rmd t'} \, \bm{S}(t) \label{22b}
    \\ &\approx \eu{\bm{\Omega}_\text{ATDC}\Delta t} \, \bm{S}(t)    
\end{align}
\end{subequations}
where in the second line we have approximated the time-ordered exponential by an ordinary exponential which is valid only for small $\Delta t$.
The final line defines the method used in practice, where
\begin{equation}
        \bm{\Omega}_\text{ATDC} = \begin{pmatrix}
            0 & -\overline{\Delta V} & 2\overline{T} \cr
            \overline{\Delta V} & 0 & 0 \cr
            -2\overline{T} & 0 & 0
        \end{pmatrix}
\end{equation}
and the integral over $\Delta V$ implied by Eq. \ref{22b} is approximated by the trapezium rule:
\begin{align}\label{DeltaV}
	\overline{\Delta V} = V_z(\bm{q}(t)) + V_z(\bm{q}(t+\Delta t))
\end{align}

This is the method which was used in our previous ab initio MASH simulation of cyclobutanone.\cite{cyclobutanone}
It is strongly based on the approach of Jain et al. \cite{Jain2016AFSSH}, who wrote the averaged time-derivative coupling in terms of a logarithm of the overlap matrix.
In turn, their approach was based on the original idea of Meek and Levine's norm-preserving interpolation. \cite{meek_evaluation_2014}
The concept behind this was effectively to approximate $T(t')$ by a linear function along the time-step.
Although this has the advantage of capturing the overall effect of a narrow avoided crossing, we see no physical reason for assuming $T(t')$ to be a linear function, especially in the vicinity of a narrow avoided crossing where it is strongly peaked.
We have therefore chosen not to name our method the ``norm-preserving interpolation'', as all the methods presented in this paper correctly preserve the norm of the spin vector, and additionally we have simply averaged the time-derivative coupling, rather than interpolating it.
We will thus refer to this method as the averaged time-derivative coupling (ATDC) approach.

\subsection{Spin propagation using local diabatization}

Plasser, Granucci et al.\cite{plasser_surface_2012}
offer a third alternative, known as the local-diabatization (LD) approach.
Here, one treats the adiabatic states as if they were the basis of a (diagonal) diabatic representation.
The propagator is therefore defined using
\begin{equation}
    \bm{\Omega}_\text{LD}(t) = \begin{pmatrix} 0 & -2 V_z(t) & 0 \\ 2 V_z(t) & 0 & 0 \\ 0 & 0 & 0 \end{pmatrix}
\end{equation}
However, to recover the correct propagation, one must take account of the fact that the basis at time $t$ is different from those at time $t+\Delta t$.
Therefore, the spin is updated by taking half of a time step in the initial basis, rotating to the new basis and finally taking half a time step in this basis:
\begin{align} \label{eq27}
    \bm{S}(t+\Delta t) = \eu{\bm{\Omega}_\text{LD}(t+\Delta t)\frac{\Delta t}{2}} \, \bm{R}_y(\chi)^\T \,\eu{\bm{\Omega}_\text{LD}(t) \frac{\Delta t}{2}} \bm{S}(t)
\end{align}
The rotation matrix 
\begin{align}
    \bm{R}_y(\chi) = \begin{pmatrix} \cos\chi & 0 & \sin\chi \\ 0 & 1 & 0 \\ -\sin\chi & 0 & \cos\chi \end{pmatrix}
\end{align}
is written in terms of the rotation angle $\chi$ as defined in Eq. \ref{ovl}.

This is almost a faithful reformulation of the method of Plasser, Granucci et al. \cite{plasser_surface_2012} from the language of wave functions into that of spin vectors.
There is however one small modification that we have introduced, to use $\eu{\bm{\Omega}_\text{LD}(t+\Delta t)\Delta t/2}\, \bm{R}_y(\chi)^\T \,\eu{\bm{\Omega}_\text{LD}(t)\Delta t/2} = \bm{R}_y(\chi)^\T \eu{ \bm{R}_y(\chi)\bm{\Omega}_\text{LD}(t+\Delta t) \bm{R}_y(\chi)^\T \Delta t/2} \,\eu{\bm{\Omega}_\text{LD}(t)\Delta t/2}$ rather than $\bm{R}_y(\chi)^\T\,\eu{[\bm{R}_y(\chi)\bm{\Omega}_\text{LD}(t+\Delta t)\bm{R}_y(\chi)^\T+\bm{\Omega}_\text{LD}(t)]\Delta t/2}$, which would correspond exactly to the original suggestion. 
Our approach 
is expected to be a slightly more accurate approximation to the time-ordered exponential of Eq.~\ref{spin ATDC}.
Even more accurate versions can be constructed by taking multiple small steps using propagators obtained by interpolating between $\bm{\Omega}_\text{LD}(t)$ and $\bm{R}_y(\chi)\bm{\Omega}_\text{LD}(t+\Delta t) \bm{R}_y(\chi)^\T$. 
In fact, an analytic solution could probably be found in terms of Weber functions, following Zener's famous derivation of an equivalent problem.\cite{Zener1932LZ}
Although these improvements may achieve a small gain in accuracy, they do not in general increase the order of the method, as the assumption of linear interpolation makes the method second order for any nonlinear Hamiltonian.

We find that the local-diabatization approach is more accurate than the approaches based on Meek and Levine's idea of linearly interpolating the time-derivative coupling 
\cite{meek_evaluation_2014,Jain2016AFSSH}.
Only the local-diabatization method can be made exact for the simple case of a Landau--Zener model, as the interpolation is perfect for a linear diabatic Hamiltonian.
The local-diabatization method is also the preferred approach in other surface-hopping implementations. \cite{Mai2018SHARC}

A simple alternative to the local-diabatization approach is available for model systems defined in terms of a global diabatic representation.
This is exploited in Runeson's simulations of exciton dynamics.\cite{Runeson2023MASH}
We will, however, focus on the standard situation for ab initio simulations of photochemistry, in which a global diabatic representation is not available.

Finally, we note that we see no need to employ multiple-time-stepping algorithms to propagate the electrons with a smaller time-step than the nuclei.
We suppose that one reason that such approaches were previously favoured is due to the use of Runge--Kutta solvers, rather than propagators based on matrix exponentials.\cite{HammesSchiffer1994FSSH,Jain2016AFSSH}
However, there is perhaps also a deeper reason, that in FSSH, the probability of hopping during an individual step must never exceed 1 and it may sometimes be necessary to use small time-steps to ensure this.
In MASH, we do not have to worry about probabilities and can take larger steps without causing such problems.

\section{
Implementation}\label{Part2}
Having discussed the various possible ways to update the nuclei and the spin vector separately, we now consider how to put these together to obtain efficient integrators for the full MASH dynamics. 
Although the equations of motion for MASH are formally time-reversible,
current ab-initio implementations of MASH\cite{Mannouch2024MASH,unSMASH,cyclobutanone} are based on algorithms originally developed for surface hopping, which are not time-reversible.\cite{Mai2018SHARC}
For instance,\\ 
    \includegraphics[width=\linewidth]{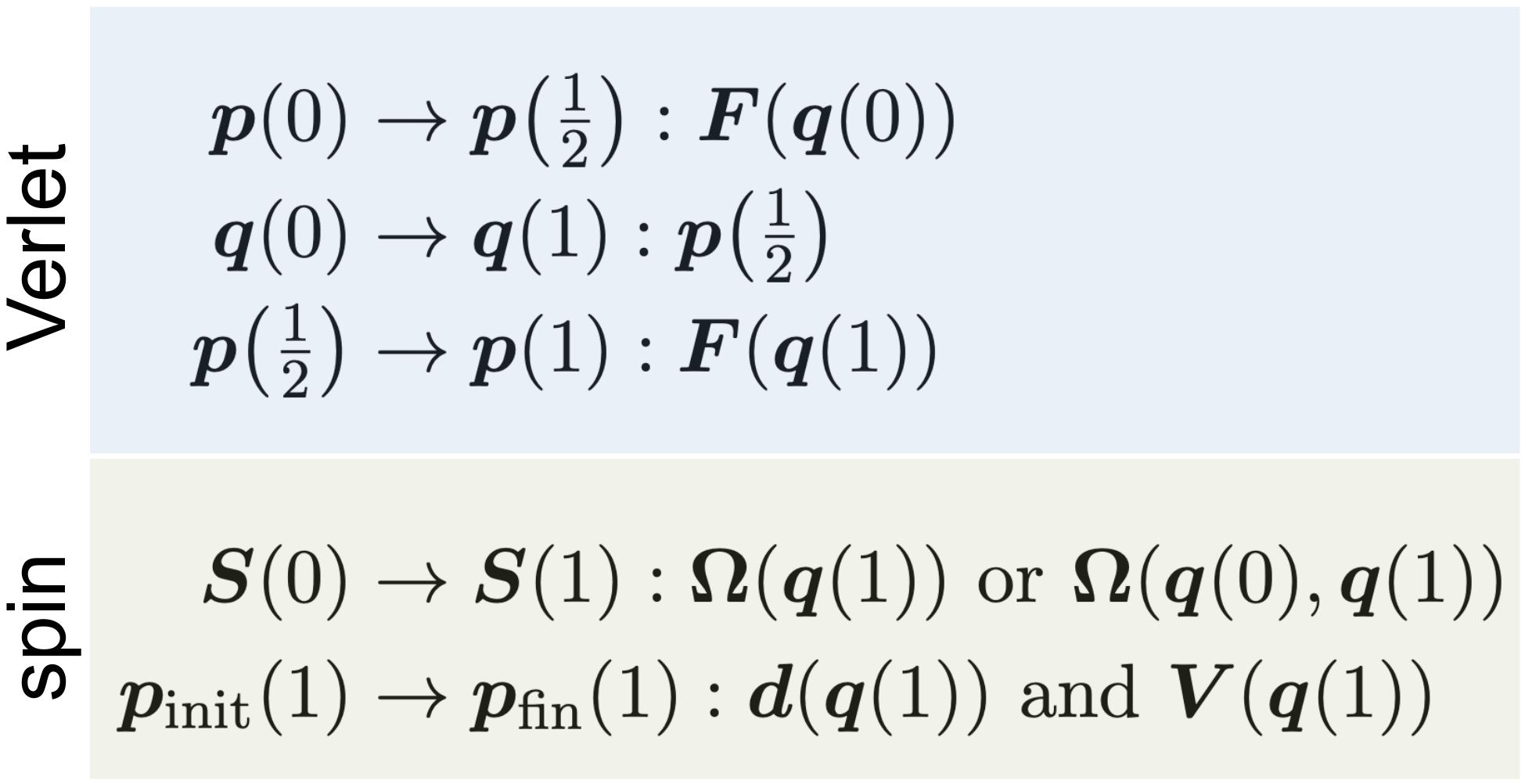}
   \label{fig:nonrev}
where 0 and 1 denote times $t$ and $t+\Delta t$ and the rest of the notation is self-explanatory.  For instance, in the first line $\bm{p}$ is propagated forwards by half a time-step according to the force measured at $\bm{q}$ at the start of the step.

This non-reversible approach can be applied using either NACs or overlaps to propagate the spin vector.
When propagating with NACs, 
there are two ways to define 
$\bm{\Omega}_\text{NAC}$, either based on information at $\bm{q}(1)$ only (which we call asym-NACs), or the more accurate approach of averaging the information at $\bm{q}(0)$ with $\bm{q}(1)$, i.e.\ $\bm{\Omega}_\text{NAC}(\bm{q}(0),\bm{q}(1))=\half[\bm{\Omega}_\text{NAC}(\bm{q}(0))+\bm{\Omega}_\text{NAC}(\bm{q}(1))]$ (which we call non-rev-NACs).
For the overlap methods called either non-rev-ATDC or non-rev-LD, the $\bm{\Omega}$ matrix is necessarily defined by both points.
In all cases, the momenta are rescaled after attempted hops in the final line, which requires the calculation of the adiabatic energies and NAC at $\bm{q}(1)$.
Overall, all these algorithms correctly reproduce MASH dynamics in the limit of $\Delta t\rightarrow0$ and some have been used in ab-initio simulations.\cite{cyclobutanone}

Nonetheless, it is desirable to develop a time-reversible algorithm, as it is nearly always a good thing to preserve a symmetry wherever possible.
In particular, in the long-time limit, it is expected that time-reversible integrators give more reliable results using larger time-steps \cite{leimkuhler_simulating_2005}. 
The trick to obtaining a time-reversible integrator is to construct it in a symmetric way such that it reads the same forwards as backwards.
A simple example for this is the velocity-Verlet algorithm which updates momenta, positions and then momenta again. 
This concept can be extended to obtain symmetric integrators for MASH, in which the phase space additionally includes the spin vector.

\subsection{Reversible NACs (rev-NACs) approach}
    To build a time-reversible algorithm using the nonadiabatic coupling vectors, the propagation of the spin $\vec S$ can be divided into two half-steps that are built around the velocity-Verlet algorithm. This is possible because knowledge of the wave functions at only one point in time is necessary to propagate $\vec S$ when using NACs. After each update of $\vec S$, we check whether a hop has occurred, in which case the momentum is rescaled or inverted. 
    \begin{figure}[H]
        \centering
        \includegraphics[width=\linewidth]{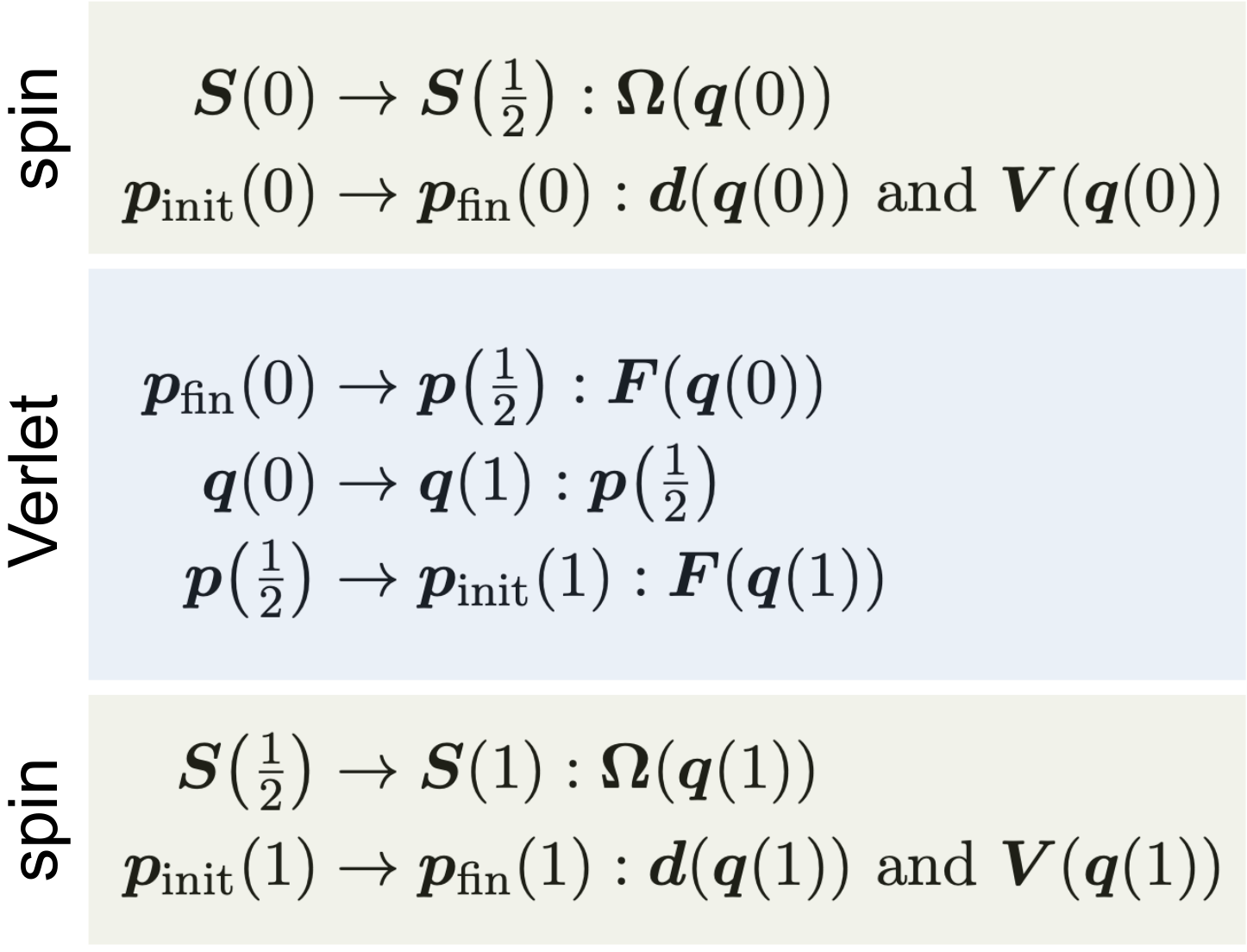}
    \label{fig:revNACS}    
    \end{figure}

Unlike the asymmetric algorithm presented above, it is clear that if you run a trajectory backwards in time using this reversible algorithm (either by using a negative value of $\Delta t$ or by flipping the signs of $\vec p$ and $S_y$), it will retrace its steps perfectly.  This is true (apart from round-off error) even for large values of $\Delta t$ and is confirmed by numerical tests.

\subsection{Piecewise-continuous time-reversible approaches}
The reversible NACs algorithm cannot be directly generalized to use overlaps.
This is because constructing the overlap requires knowledge of the wave functions at two points in time. It is therefore not possible to update the spin before the nuclei.
It is thus necessary completely to restructure the algorithm when using overlaps.

To proceed, we first notice that the standard algorithms (non-rev-NACs, non-rev-ATDC, non-rev-LD) are actually time-reversible for any step in which no hops occur. This is because during these steps, the nuclear variables and the spin vector are not coupled and the propagators for these two separate parts are themselves reversible. 
In particular, the nuclei are propagated using the time-reversible velocity-Verlet algorithm.
In order to prove that the spin update is reversible, we need to demonstrate that the propagator which takes $\bm{S}(t)$ to $\bm{S}(t+\Delta t)$ is equal to its adjoint. \cite{leimkuhler_simulating_2005} The adjoint of a propagator $\mathcal{U}(\Delta t)$ is defined as  $[\mathcal{U}(-\Delta t)]^{-1}$. 
In other words, propagating backwards in time should be the inverse of forward propagation.
This is easy to show for non-rev-NACs, where $\bm{\Omega}_\text{NAC}$ is defined as the average of the information from the initial and final points.
Similarly, non-rev-ATDC makes use of time-averages, \eqn{DeltaV} and \eqn{Tbar}, which are symmetric to time reversal, i.e.\ $\bm{\Omega}(t,t+\Delta t)=\bm{\Omega}(t+\Delta t,t)$
such that $\eu{\bm{\Omega}(t,t+\Delta t) \Delta t} = (\eu{-\bm{\Omega}(t+\Delta t,t)\Delta t})^{-1}$.
Finally, it can be shown that the inverse of the propagator of non-rev-LD, \eqn{eq27}, obeys:
\begin{multline}
    \eu{\bm{\Omega}_\text{LD}(t+\Delta t)\frac{\Delta t}{2}} \bm{R}_y(\chi(t,t+\Delta t))^\T \,\eu{\bm{\Omega}_\text{LD}(t) \frac{\Delta t}{2}}
    \\= \left(\eu{-\bm{\Omega}_\text{LD}(t)\frac{\Delta t}{2}} \bm{R}_y(\chi(t+\Delta t,t)^\T \,\eu{-\bm{\Omega}_\text{LD}(t+\Delta t) \frac{\Delta t}{2}}\right)^{-1}
\end{multline}
This follows because $(\bm{R}_y(\chi(t+\Delta t,t)^\T)^{-1}=\bm{R}_y(\chi(t,t+\Delta t))$ due to the properties of orthogonal rotation matrices and because ${\chi(t,t+\Delta t)}=-\chi(t+\Delta t,t)$ due to the antisymmetry of the overlap matrix. 
Taken together, these arguments demonstrate that the spin is evolved in a time-reversible fashion by all these methods, i.e.\ propagating forwards and then backwards will return the spin exactly to its initial value.

When the sign of $S_z$ changes within a time-step, however, this is only noticed by the non-reversible algorithm at the end of the step. As a result, the nuclei are propagated on a single electronic surface for the full step. When this step is reversed, the nuclei again evolve on a single surface, but this time on the other surface, causing a mismatch between forward and backward trajectories.

We will therefore use the standard algorithm as our building block and adjust it in a way to ensure time-reversibility even in steps that include hops.
This is achieved by dividing the step into two parts at the time of hopping $\tau$.
The nuclei and spin are first propagated until $\tau$, using the standard algorithm. Then, the hop is performed and the momentum is rescaled before the standard algorithm is applied again to propagate to the end of the step: 
\begin{figure}[H]
    \centering
    \includegraphics[width=\linewidth]{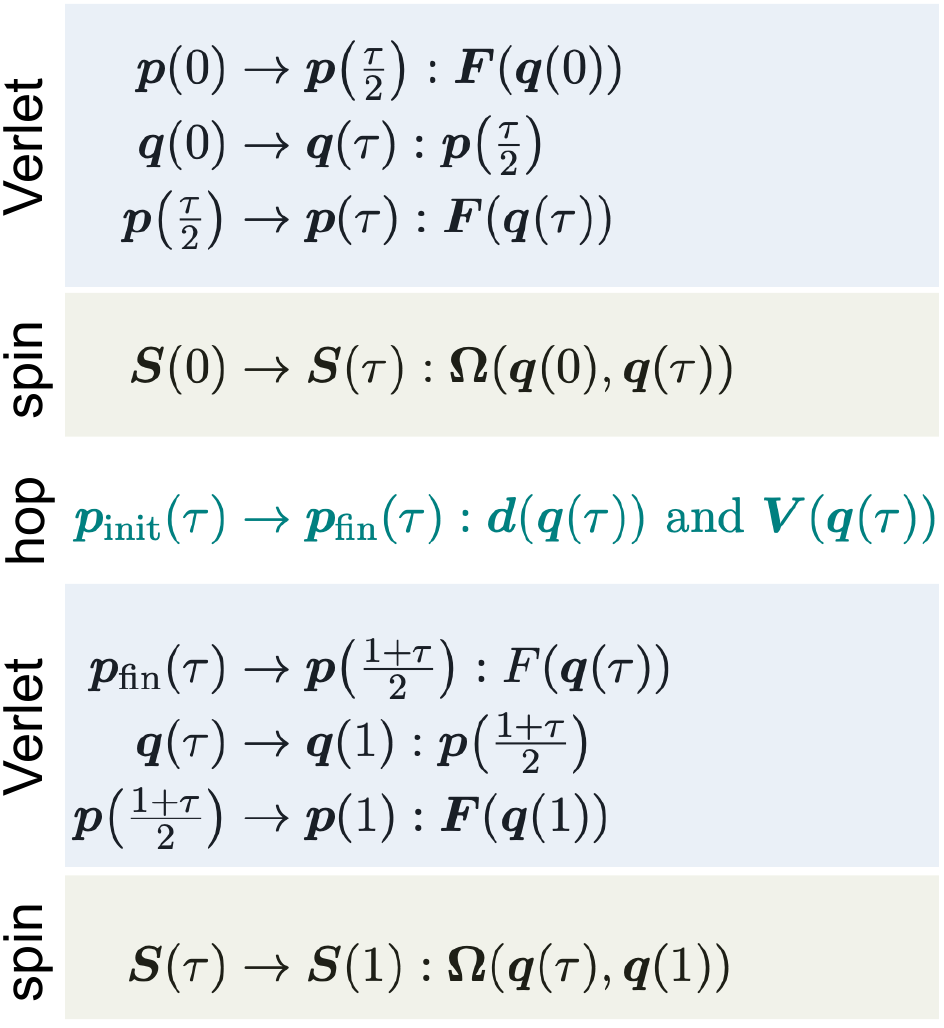}
    \label{fig:ovl}
\end{figure}
Note that only one rescaling step is needed as we assume that maximally one hop can occur during a full time-step.
Numerical tests show perfect agreement between the forwards and backwards propagation, confirming that the algorithm is time-reversible.
We call these methods piecewise-continuous (pc) as the discontinuity introduced by the hop is carefully controlled between two continuous evolutions.
Various spin-propagation approaches can be used, leading to the methods rev-pc-NACs, rev-pc-ATDC and rev-pc-LD.
To ensure that rev-pc-NACs is reversible, it is necessary to average the NACs between 0 and $\tau$ or $\tau$ and 1 as appropriate, similarly to non-rev-NACs.

\begin{figure} [H]
    \centering
    \includegraphics[width=\linewidth]{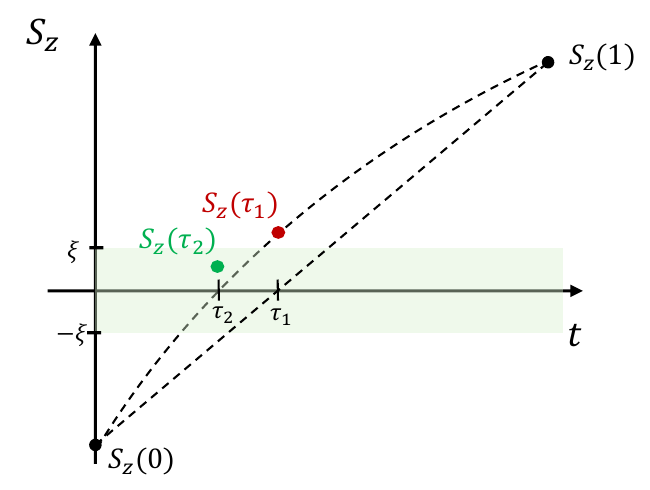}
    \caption{Example of a root search for the hopping time, $\tau$; after a state change has been detected, the first estimate is $\tau_1$, but $S_z(\tau_1)$ does not lie within the error band. Thus, a new estimate $\tau_2$ is made using a spline interpolation between  $S_z(0)$, $S_z(\tau_1)$ and $S_z(1)$. This fulfills the condition given by $\xi$ and the search is finished.}
    \label{fig:interpolation}
\end{figure}

These methods of course require that one knows the time, $\tau$, at which the hop will occur.
This point can be found using any one-dimensional root search.  In this work, we employed the following algorithm: First, the propagation is performed for the full time-step and the state before and after the step are compared to check for hops. If there was no change, the step is finished. If there is 
an attempted hop, an estimate of $\tau$ (called $\tau_1$) is made using a linear interpolation between $S_z(0)$ and $S_z(1)$.  The propagation is then recalculated from $t = 0$ to $t = \tau_1$ using the steps as shown above. If $S_z(\tau_1)$ now lies within an error band around the equator of the Bloch sphere specified by the parameter $\xi$, $\bm{p}$ is rescaled and the variables are propagated to $t = 1$ to finish the step. If $S_z(\tau_1)$ does not fall into this region, a better estimate for $\tau$ (called $\tau_2$) is calculated using spline interpolation between all values of $S_z$ that are obtained so far.
This procedure is iterated until convergence (see Fig. \ref{fig:interpolation}).
It is important to emphasize that there are only two steps in the final propagation, divided at the converged value of $\tau$.

Overall, the algorithm becomes rigorously time-reversible in the limit that $\xi\rightarrow0$.
We use a value of $\xi$ of $10^{-4}$ and find that this typically requires only about 2--3 iterations to obtain a sufficiently converged value of $\tau$. 
Numerical tests confirm that trajectories retrace their steps almost perfectly when time is run backwards.
The piecewise-continuous methods are formally implicit, meaning that an iterative approach is required.
Implicit approaches are not commonly favoured in molecular science as they increase the overall cost of the simulation \cite{leimkuhler_simulating_2005}.
However, in our case, the iteration is only required on steps which hop and as hopping events are rare on the timescale of a full trajectory, the increase in computational cost from these iterations is not noticeable. 
As we shall show, the benefits far outweigh the extra complexities.

\section{Results and discussion}
This paper has introduced new time-reversible methods, based either on nonadiabatic coupling vectors (rev-NACs and rev-pc-NACs) or  wave-function overlaps (rev-pc-ATDC and rev-pc-LD). 
We shall compare these with the original non-reversible algorithms (asym-NACs, non-rev-NACs, non-rev-ATDC, non-rev-LD) to illustrate the differences in accuracy for various choices of time-step, $\Delta t$.

To illustrate the reversibility of the new methods we analysed individual trajectories ran forwards and
then backwards in time. Moreover, we implemented
variable time-stepping which is only possible
due to the deterministic nature of the theory.
This allows to set a large time-step for the simulation which is divided into smaller steps
whenever the large time-step results in inaccuracies. This requires that a condition is checked
after each step to determine whether significant errors were made. For instance, as the
velocity-Verlet algorithm itself is not exactly energy conserving and thus, the conservation of energy
can be a useful indicator for inaccurate propagation due to a too large time-step.

\begin{figure}
    \centering
    \includegraphics[width=\linewidth]{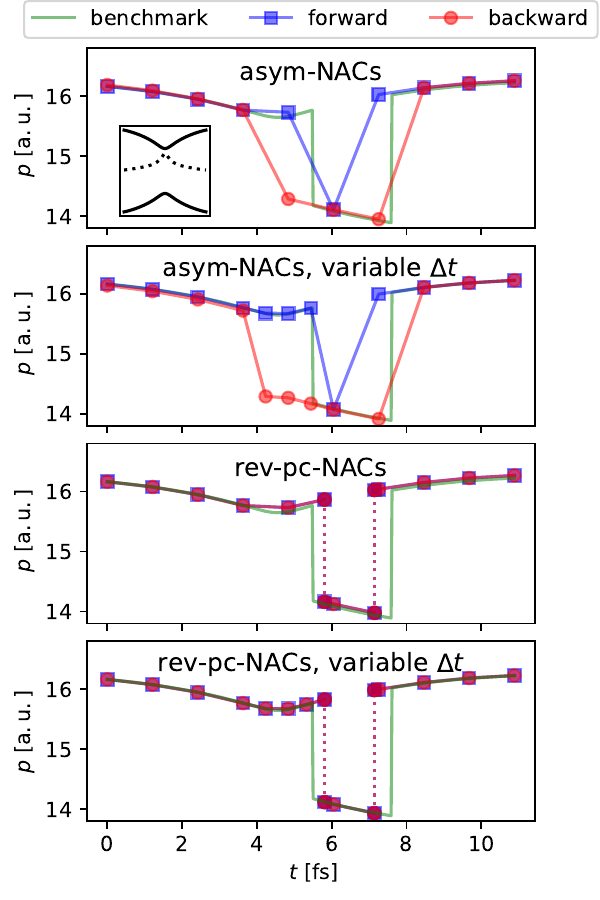}
    \caption{Example trajectory for the propagation forwards and backwards in time on the Tully model with initial conditions $q(0) = -1.5,$ $p(0) = 16.16$, and the normalized vector of $\bm{S}(0) = [0.02,0.056,-0.998]$. 
     The potential is shown in an inset with the PES in solid and the NAC in dotted lines, scaled by a factor of 1/400. 
    Each method is compared to a benchmark trajectory computed at a smaller time-step where all methods agree.
    The variable time-stepping algorithm recursively bisects steps in which $\Delta \mathcal{E} \ge 10^{-4}$. In the plot, the size of the time-steps taken is indicated by the spacing of the dots.
    }
    \label{fig:fwdbwd}
\end{figure}

In Fig \ref{fig:fwdbwd}, we show an example trajectory propagated first forwards and
then backwards in time on Tully's simple model of an avoided crossing \cite{tully_molecular_1990} using asym-NACs and rev-pc-NACs, with and without variable time-stepping. One can see that the rev-pc-NACs method retraces its steps perfectly, while the asym-NACs method shows a large difference between the forward and backward trajectories, which is not completely resolved even after both trajectories hop twice.
The situation is even worse in other cases where the number of hops differs (not shown).

Using variable time-stepping 
yields only minimal improvements but 
does not resolve the discrepancy between forward and backward asym-NACs trajectories
and does not correct the small error in the hopping time of rev-pc-NACs. 
Overall, these results indicate that it is not particularly helpful to use variable time-stepping in this case.
It is, however, clear that more accurate simulations are obtained using reversible integrators than variable time-stepping with the standard approaches.

\subsection{Error analysis}

When we discuss the error made by an integrator, we must distinguish between the truncation (local) error of a single step and the global error at a specified time after many steps.
The concatenation of local errors typically causes the global error to appear at a lower order of $\Delta t$.
For example, the velocity-Verlet algorithm has $\mathcal{O}(\Delta t^3)$ truncation error. 
However, it is known as a second-order method as its global error is $\mathcal{O}(\Delta t^2)$. \cite{leimkuhler_simulating_2005}
This is justified by a simple argument. Using $\mathcal{U}$ to represent the propagator (known as a dynamical map): 
$\mathcal{U}_\text{approx}(t_\text{max}) = [\mathcal{U}_\text{exact}(\Delta t) + \mathcal{O}(\Delta t^3)]^N$, where $N=t_\text{max}/\Delta t$ is the number of time-steps used to propagate to time $t_\text{max}$.
Expanding using the binomial series gives $\mathcal{U}_\text{approx}(t_\text{max}) = \mathcal{U}_\text{exact}(\Delta t)^N + \mathcal{O}(N\Delta t^3) = \mathcal{U}_\text{exact}(t_\text{max}) + \mathcal{O}(\Delta t^2)$, as we wished to show.

The simple reason for the excellent behaviour of the velocity-Verlet algorithm (relative to alternative approaches such as the Euler method) is due to its time-reversible nature.
This symmetry ensures that 
the order of the method is even.
In contrast, non-reversible algorithms may have odd order.
Therefore, unless one works hard to construct a complicated high-order algorithm, the global error will typically be $\mathcal{O}(\Delta t)$, i.e.\ a first-order method.

A formal error analysis could of course be carried out mathematically for our proposed algorithms.
However, this is complicated by the fact that MASH trajectories are not continuous, let alone twice differentiable as is required by textbook derivations.\cite{leimkuhler_simulating_2005}
It is therefore much simpler to present illustrative examples.

We  again employed the Tully model of an avoided crossing and read off the order of error from the numerical data.
We consider results from both single-trajectory calculations as well as averages over many trajectories.
In the latter case, the initial spin vector $\vec S$ was sampled uniformly from the lower half of the Bloch sphere (with unit radius) and the nuclear coordinates were sampled from the Wigner distribution 
\begin{equation}
    \rho_\mathrm{W}(p, q) = 2 \text{ exp}\left(- \frac{1}{\gamma} \, (p-\Bar{p})^2 - \gamma(q - \Bar{q})^2\right)
\end{equation}

The order of the global error can be obtained numerically from the slope of a logarithmic plot of the error vs $\Delta t$.
In this work, we define the global error for a trajectory as
\begin{align}\label{error}
     \frac{1}{N_\mathrm{steps}+1}\, \sum_{n=0}^{N_\mathrm{steps}} \big\lVert x_\mathrm{exact}(n\Delta t)-x_\mathrm{approx}(n\Delta t)\big\rVert
\end{align}
where $x$ can be any of the phase-space variables (position, momentum, or spin) and $ x_\mathrm{exact}$ is a benchmark trajectory run at a very small time-step where all of the methods agree.

\begin{figure} 
    \centering 
    \includegraphics[trim={0mm 0mm 0mm 0mm}, clip,   width=\linewidth]{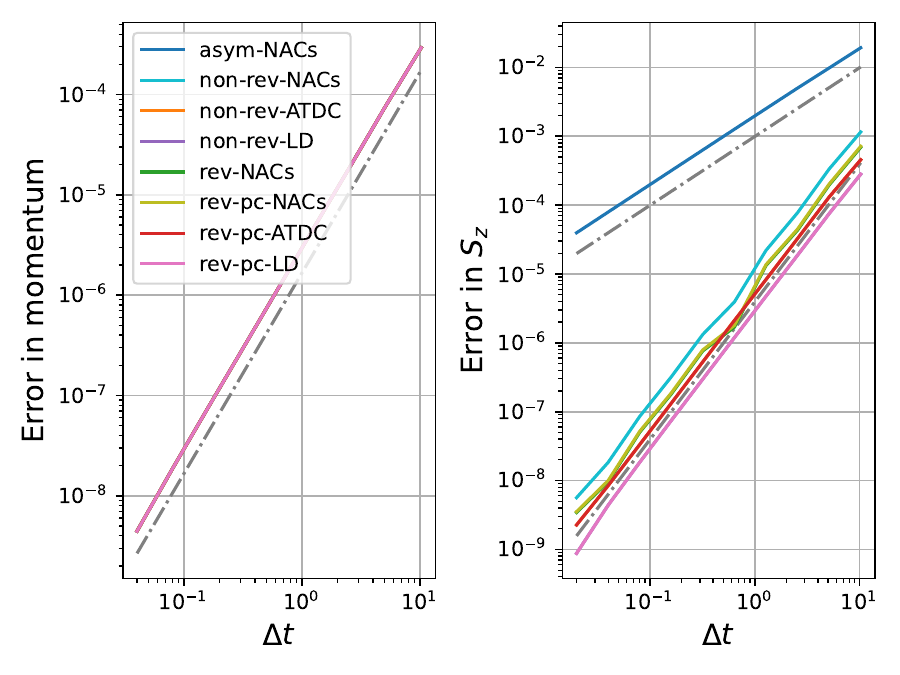} 
    \caption{Comparison of the order of the global error for the different implementations of MASH for a generic trajectory without hops going from left to right ($q_0 = -1.5, \, \, p_0 = 15.6$, $ \vec S =[-0.04,-0.08,0.9]$) in the Tully model.
    All methods show a global error of order two for the momentum due to the underlying velocity-Verlet algorithm (all coloured lines are on top of each other in the left graph). However, only the symmetric methods are of order two for $S_z$. The dashed grey lines have slopes of 1 and 2 for comparison.}
    \label{fig:order}
\end{figure}

We first look at the case of a trajectory in which no hops occur.
It is important to notice that in this case, position and momentum are updated exclusively by the velocity-Verlet algorithm, which is second order.
This is illustrated clearly by the second-order dependence of the error in the momentum on the time-step as shown in \fig{order}(a).
However, the methods differ in the way they update the spin vector.
In particular, when one looks at the $S_z$ variable, shown in \fig{order}(b), it is clear that the asym-NACs algorithm is only first order, whereas all the other (symmetric) algorithms are second order.
Note that in this case it is not the NACs method \emph{per se} that causes the poor performance, but rather the asymmetry of the spin-propagation algorithm. 
This example shows clearly how much higher accuracy can be achieved using symmetric algorithms for the same choice of time-step.

Additionally, we note that in this case the LD method is slightly more accurate than the ATDC or NACs approaches.
In cases where the avoided crossing is even sharper (not shown), the NACs method performs significantly worse than the two methods based on overlaps, as expected from previous work. \cite{meek_evaluation_2014, plasser_surface_2012}

\begin{figure} 
    \centering 
    \includegraphics[trim={0mm 0mm 0mm 0mm}, clip,   width=\linewidth]{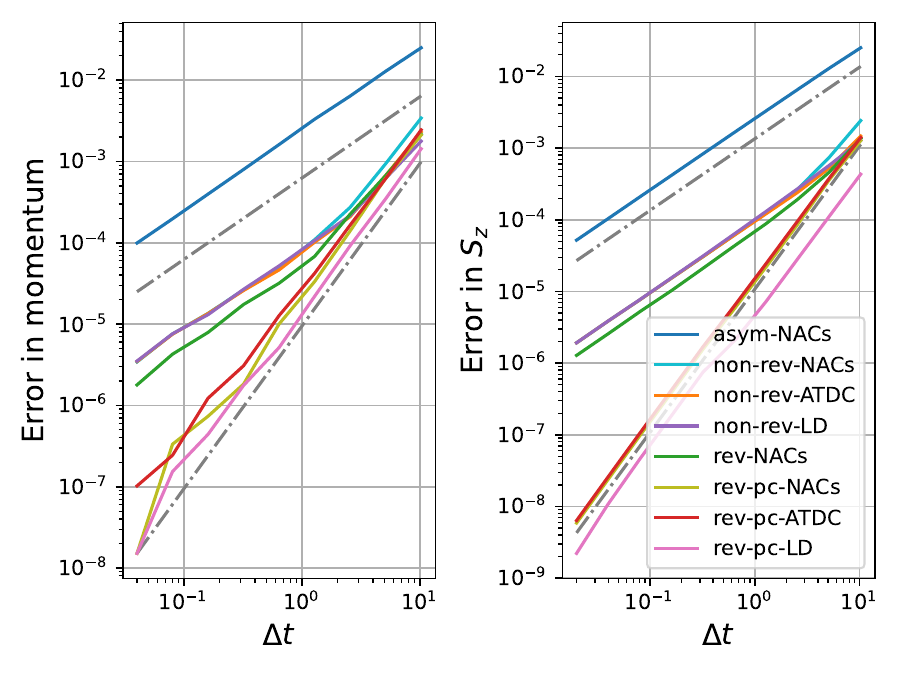} 
    \caption{Comparison of the order of the global error for the different implementations of MASH averaged over 5000 trajectories sampled from a Wigner distribution centred around $\Bar{q}=-1.5$ and $\Bar{p}=\sqrt{0.2m}$ with $\gamma = 0.1$. Some trajectories do not hop, others hop once, twice or more, with an average of about one hop. Only the piecewise-continuous methods preserve the second-order accuracy while the rev-NACs and non-reversible methods become first order due to the $\mathcal{O}({\Delta t})$ error made during hops.
    }
    \label{fig:order3}
\end{figure}

The same process was repeated for an ensemble of trajectories which included hops.
To ensure a fair comparison, exactly the same set of initial conditions was used for each method.
The results are shown in \fig{order3}.

The first thing to notice is that due to the coupling between nuclear and electronic degrees of freedom, even the error in momentum becomes first order for the non-rev-NACs method.
We also see that other non-reversible methods (non-rev-ATDC and non-rev-LD) are first order.
It is perhaps surprising at first sight that the rev-NACs method also appears to be first order, even though it is a time-reversible method.
This observation is explained by noting that the textbook global-error analysis does not directly apply to MASH algorithms as the momenta are not continuous across a hop.
Each time a hop is encountered by these methods, an error of $\mathcal{O}(\Delta t)$ is made.  Assuming there are only a finite number of hops along a trajectory implies that the global error will also be of $\mathcal{O}(\Delta t)$, as confirmed in the results.

The only methods which remain second order are the reversible piecewise-continuous algorithms. 
These methods avoid the difficulty of the discontinuity by dividing the propagation into two continuous steps, as the name suggests. It does not introduce an $\mathcal{O}(\Delta t)$ error when hops are performed and thus preserves the second-order nature of the methods.
This results in an increased accuracy of several orders of magnitude.

\subsection{Example application to pyrazine}
After formally analysing the order of the error, we next test the implementations on a more relevant chemical problem in multiple dimensions. 
For this, we chose to simulate the internal conversion in a 2-state 3-mode linear-vibronic-coupling model of pyrazine \cite{schneider_s1-s2_1988}.
We initialized the simulations from a harmonic Wigner function in the higher diabatic state and measured the decay of this diabatic population.
Diabatic observables are obtained according to the standard MASH procedure as linear combinations of correlation functions of adiabatic populations and coherences including the appropriate weighting factors.\cite{mannouch_mapping_2023,richardson_nonadiabatic_2025}
A comparison with exact quantum-mechanical results is presented in previous work, showing that MASH gives a reasonable approximation.\cite{mannouch_mapping_2023}
Here, we focus only on the numerical accuracy of the implementation.

\begin{figure*}
    \centering
    \includegraphics[trim={0mm 0mm 0mm 0mm}, clip,  width=\textwidth]{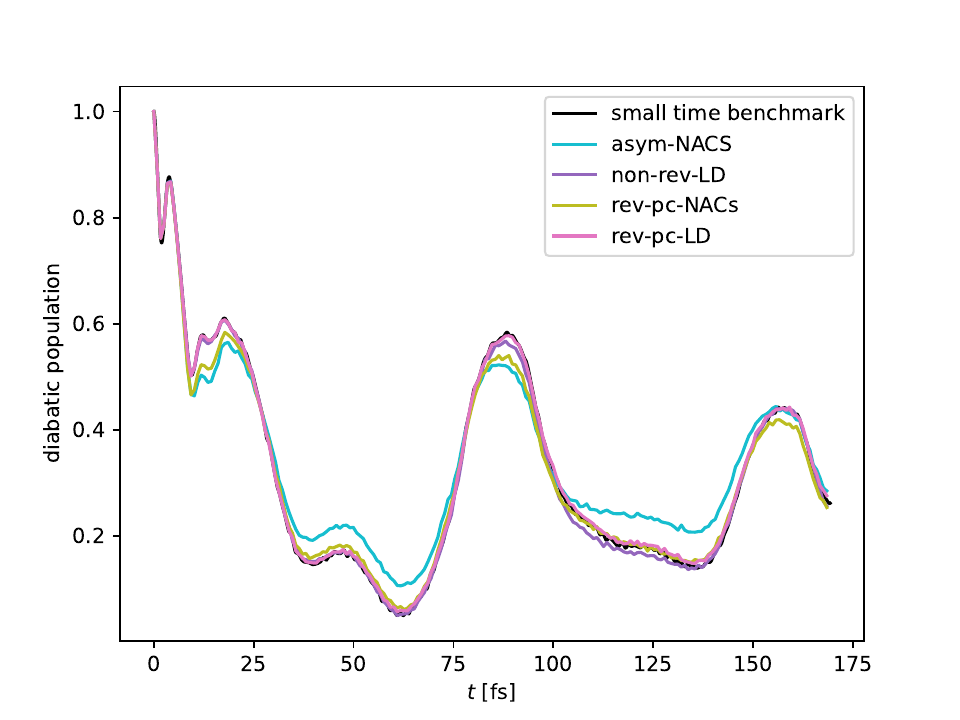}
    \caption{Comparison of the time-dependent diabatic population for a 2-state 3-mode model of pyrazine computed with $\Delta t =1.2$ fs to a benchmark obtained with $\Delta t_0 = \Delta t/350$, where all methods agreed. The reversible, piecewise-continuous methods outperform their non-reversible counterparts. Additionally, the LD methods based on overlaps perform significantly better than those based on NACs. ATDC (not shown) is similar to LD in this case} 
    \label{fig:pyrazine}
\end{figure*}

To test the accuracy of the various algorithms, a benchmark result was obtained as the average of 100,000 trajectories with a very small time-step, $\Delta t_0$, where agreement was seen between all methods. 
The calculation of the trajectories was then repeated with the same set of initial conditions for a large time-step $\Delta t$. 
As shown in Fig.~\ref{fig:pyrazine} the reversible, piecewise-continuous version of each algorithm is more accurate than its non-reversible counterpart. Moreover, the rev-pc-LD algorithm performs best and shows almost perfect agreement with the benchmark, whereas the non-rev-LD deviates in the regions of $80-90$ and $100-140$ fs. Both NACs methods (and especially asym-NACs) perform significantly worse, which is expected as the well known disadvantage of using NACs is that they can be strongly peaked for systems with weak diabatic couplings such as in the vicinity of a conical intersection \cite{meek_evaluation_2014}. This means that a short time-step may be necessary to reproduce the correct dynamics. Overall, these results show that the new reversible implementations of MASH allow for a larger time-step with no loss in accuracy.  

Moreover, we also employed variable time-stepping but found that it does not noticeably improve the accuracy of the simulations. Instead, even with a low threshold for the maximum allowable change in the energy, the performance showed no noticeable improvement, despite incurring three times the computational cost.
This implies that one should look for a better indicator than energy conservation to detect when smaller time-steps are required. For example, one could use the smaller time-steps only when hops are encountered. However, we have shown that variable time-stepping is not necessary when using the new reversible, piecewise-continuous methods.
Additionally, locating the exact time of hopping is more efficient using interpolation than bisection\cite{Runeson2023MASH} traditionally used in the variable time-stepping approach.

\section{Conclusion}
This work has enhanced the mapping approach to surface hopping (MASH) by developing time-reversible, second-order integrators. This improves the accuracy of the trajectories for a given finite step-size, allowing larger time-steps to be made for a given error tolerance. 



Important to note is that constructing a time-reversible integrator alone is not sufficient to ensure second-order accuracy. This subtle point arises because of the discontinuity of MASH trajectories when the electronic state is changed, which introduces an error of $\mathcal{O}(\Delta t)$ whenever a hop occurs. This lowers the order of the global error of rev-NACs to first order.
In order to achieve second-order accuracy it was necessary to develop piecewise-continuous integrators by locating the exact time of hopping to a high accuracy. 
The methods differ in the way in which the spin is propagated; this can be based on NACs (rev-pc-NACs) or overlaps (rev-pc-ATDC and rev-pc-LD).
These exhibit a smaller global error of the numerical integrator, $\mathcal{O}(\Delta t^2)$, compared to the previous $\mathcal{O}(\Delta t)$. 
This leads to a marked improvement in accuracy when calculating correlation functions and observables with large time-steps. 

Additionally, there may be particular advantages in using time-reversible integrators when employing rare-event theories, such as transition-path sampling\cite{TPSreview}. Time-reversal symmetry can also simplify the procedure of obtaining reaction rates.\cite{MASHrates}

These enhancements demonstrate a distinct advantage of MASH over methods like fewest switches surface hopping (FSSH), which inherently cannot employ time-reversible integrators due to their stochastic nature.
This suggests that MASH can employ a larger time-step than FSSH, making the method a more robust and efficient method for nonadiabatic molecular dynamics, in addition to the higher intrinsic accuracy exhibited in previous studies.

\begin{acknowledgement}
J.A.G. thanks the German Academic Scholarship Foundation which supports her undergraduate studies.
\end{acknowledgement}

\bibliography{references, amirasreferences}

\end{document}